# Bihar Assembly Elections 2020: An analysis

Mudit Kapoor[1] and Shamika Ravi[2]

**Abstract**

We analyze the Bihar assembly elections 2020, and find that poverty was the key driving factor, over and above female voters as determinants. The results show that the poor were more likely to support the NDA. The relevance of this result for an election held in the midst of a pandemic, is very crucial, given that the poor were the hardest hit. Secondly, in contrast to conventional commentary, the empirical results show that the 'AIMIM factor' and the 'LJP factor' hurt the NDA while benefitting the MGB, with their presence in these elections. The methodological novelty in this paper is combining elections data with wealth index data to study the effect of poverty on elections outcomes.

*JEL Keywords*: poverty, elections, Bihar

1. Introduction

Bihar assembly elections were an important litmus test for the policies of the National Democratic Alliance (NDA)[3] particularly in the aftermath of the socio-economic upheaval caused by COVID–19. Unfortunately, the poor have had to face a disproportionately higher brunt of the crisis. The results in favor of the NDA came as a surprise (it was counter to what majority of the exit polls had predicted) primarily due to the following factors, first and foremost, the anti-incumbency; second, the humanitarian and economic crisis precipitated by COVID–19; and third, to a limited extent, a spirited campaign, centered around jobs and unemployment, led by the leader of Rahstriya Janta Dal (RJD), the leading party of the Maha Gath Bandan (MGB)[4]. Political experts while attempting to explain NDA's surprising victory over MGB have focused primarily on two classes of voters; (i) women voters, who recalling the misrule of the RJD, favored the NDA, and (ii) the Muslim vote which was diverted away from the MGB by parties such as All India Majlis-E-Ittehadul Muslimeen (AIMIM).

---

[1] Associate Professor, Economics, ISI, Delhi, mudit.kapoor@gmail.com
[2] Non-resident Senior Fellow, Brookings Institution, sravi@brookings.edu
[3] NDA alliance included Janta dal (United) (JDU), Bharatiya Janta Party (BJP), Vikassheel Insaan Party (VIP), and Hindustani Awam Morcha (Secular) (HAMS).
[4] MGB alliance included Rashtriya Janata Dal (RJD), Indian National Congress (INC), Communist Party of India (CPI), Communist Party of India (Marxist) (CPM), and Communist Party of India (Marxist-Leninist) (Liberation) (CPIML).

In this paper we perform an exploratory data analysis to isolate factors that have determined outcome in these elections. To our knowledge, this is the first paper to explore relationship between poverty and election outcomes. Methodologically the innovation in the study is to combine datasets across different sources for detailed analysis. In particular, we look at the following factors as determinants of election outcomes:

(a) Sex ratio of the electorate (female per 1000 male electors): this is an important determinant given earlier studies which have established the significant role of female voters as agents of change in Bihar elections.[5]

(b) Proportion of the poorest households within a state in the district in which the constituency is located: these elections were held in the midst of a pandemic which has seen greatest suffering among the poorest people in the population, we want to isolate the relationship between poverty and election outcomes.

(c) Proportion of general caste households in the district in which the constituency is located: given the significant role of caste in Indian elections, this is an obvious line of enquiry for Bihar elections.

(d) Proportion of Muslim population in the district in which the constituency is located: this is an important line of enquiry given the belief that Muslims are a distinct electoral demography.[6]

(e) AIMIM factor – we analyze this to identify the effect it might have on performance of other political parties in these elections.

(f) LJP factor – we analyze this to identify the effect it might have on performance of other political parties in these elections.

---

[5] 'Women voters can tip the scales in Bihar', The Hindu: https://www.thehindu.com/opinion/op-ed/women-voters-can-tip-the-scales-in-bihar/article7470851.ece

[6] Maqbool Ahmed Siraj (1986) Electoral demography of Indian Muslims, Institute of Muslim Minority Affairs. Journal, 7:2, 557-603, DOI: 10.1080/13602008608716003

## 2. Data Source and Covariates

A novelty in this paper is that we combine data from different sources to analyze election outcomes. In particular, we combine data from Election Commission of India (ECI) and National Family Health Survey (NFHS). Our main source of data on 243 assembly constituencies in the Bihar general assembly election in 2020 is from the ECI which provides data at the assembly constituency level on the candidates that contested, the number of votes secured by each candidate, and the political party each candidate is affiliated with. If a candidate is not affiliated to a political party then the candidate is classified as independent. In addition to the election results we use data from the 2015 Bihar general assembly election from the ECI to compute the sex ratio of the electorate (female per 1000 male electors) at the constituency level. We then classify the constituencies into three groups, (i.e. sex ratio of the electorate, <861, 861 to 886, and ≥886).

Next, we use household data from the National Family Health Survey, round 4 (NFHS IV), which was conducted in 2015–2016. The advantage of the NFHS IV dataset was that the sample size was large enough to construct socio-economic indicators at the district level. First, we construct data on proportion of poorest households in a district using the wealth index. The wealth index was constructed by assigning a score to each household, which was based on ownership of consumer goods such as television, bicycle, car, etc., characteristics of house, such as source of drinking water, toilet facilities, material used for floors, and walls, etc. The scores are assigned using a principal component analysis. Households are then ranked based on the score and divided into five equal categories (quintiles), and the households in the lowest quintile were classified as the poorest. Then for each of the 38 districts in Bihar we construct the proportion of poorest households. Next, we classify districts intro three equal groups based on the proportion of the poorest households (<15.5%, 15.5% to 25.5%, ≥25.5%).

Second, the NFHS IV household data also identifies the religion and the caste (which are classified into other backward class, scheduled caste, and scheduled tribe) of the head of household. If the household head states her or his religion as Muslim then the household is classified as a Muslim household. We classify the household as belonging to general category if the head of the household does not belong to either of the caste category (other backward class, scheduled caste, scheduled tribe). Then for each district we construct the proportion of Muslim households, and classify them into three equal groups (i.e. proportion of Muslim households, <8.6%, 8.6% to 15.33%, ≥15.33%). Similarly, we classify each district into three equal groups by the proportion of general category households (i.e. proportion of general category households, <14.5%, 14.5% to 19.6%, ≥19.6%). Third, we classify districts into three equal groups based on proportion of households residing in rural areas (i.e., <86.6%, 86.6% to 92.3%, ≥92.3%).

We then classify each constituency as belonging to one of the groups in terms of proportion of poorest households, proportion of Muslim and general caste households, and proportion of households residing in rural areas, depending on the districts in which the constituency is located. In addition, we use the data from the ECI, to construct a dummy variable which takes a value one if AIMIM contested the election in that constituency and zero otherwise, similarly we construct a dummy variable which takes a value one if LJP contested the election in that constituency and zero otherwise.

3. Results

The first set of results – summary statistics - are reported in table 1. There were 243 assembly constituencies, where elections were held in Bihar. The contest was primarily between the NDA[7] and the MGB[8]. In terms of the total seats the NDA won in 125 out of the total 243 assembly constituencies while the MGB won 110 assembly constituencies. However, in terms of the total votes, the NDA received 15,701,226 (37.3%) out of the total votes 42,137,620 that were cast, while the MGB received 15,688,458 (37.2%) of the total votes, a difference of 12,768 votes.

The results suggest a high degree of variance in the performance of individual political parties. For example, the BJP of the NDA had the best strike rate (seats won as a proportion of seats contested) of 67.3%, while INC of the MGB had the worst strike rate of 27.1%. In terms of the percentage of votes received by a political party across the constituencies that they contested in, the BJP received 42.6% of the votes while RJD received 38.9% of the votes.

---

[7] (NDA alliance included Janta dal (United) (JDU), Bharatiya Janta Party (BJP), Vikassheel Insaan Party (VIP), and Hindustani Awam Morcha (Secular) (HAMS))
[8] (MGB alliance included Rashtriya Janata Dal (RJD), Indian National Congress (INC), Communist Party of India (CPI), Communist Party of India (Marxist) (CPM), and Communist Party of India (Marxist-Leninist) (Liberation) (CPI (ML) (L))).

For our next set of results, we construct three variables at the level of the alliance (NDA and MGB) and the political parties (JDU, BJP, RJD, and INC): (a) total seats contested, (b) total seats won, and (c) strike rate, which is the ratio of seats won to seats contested and relate their performance to the factors considered (see table 2b). For example, in 81 assembly constituencies that NDA contested in, the sex ratio of the electorate was < 861 female per 1000 male electors, and NDA won in 36 of those constituencies with the strike rate of 44.4%. Similarly, the NDA contested in 81 constituencies that had sex ratio of the electorate ≥886 and won in 52 of them with a strike rate of 64.2%. If we look at the performance of INC in the constituencies where they contested and the AIMIM also contested, there were 7 such constituencies, INC won 4 of them with the strike rate of 57.3%, while in other 63 constituencies in which the INC contested where AIMIM did not contest, it won 15 seats with the strike rate of 23.8%.

To sharpen our analysis, we perform a logistic regression, where primary outcome of interest was winning in a constituency in which the political party (and alliance) contested the election. In particular, we ran the following regression:

$$logit(\pi_i) = constant + Covariates_d + AIMIM\ factor + LJP\ factor + error\ term_i,$$

where subscript *i* denotes the constituency, *Covariates_d* are the factors at the district level *d* where the constituency is located, these covariates have been explained in detail in the data section. *AIMIM factor* is a dummy variable which takes a value one if AIMIM contested the election in the constituency and zero otherwise, and *LJP factor* is a dummy variable which takes a value one if LJP contested the election in the constituency and zero otherwise, and *error term* are the error terms that are clustered at the district level to account for similarities in the constituencies within a district. The summary data of the covariates are presented in table 2a. The results of the logistic regression are presented in table 3, figure 1a, and figure 1b. These results are presented in the form of Odds Ratios – an odds ratio measures the association between an outcome and an exposure.

Here the outcome of interest is a party/alliance winning and the exposures are various factors that have been considered.

We find that NDA had 4.25 times higher odds of winning an election in constituencies that were located in the poorest district (where proportion of poorest households was ≥25.6%) as compared to wealthier districts (where proportion of poorest was <15.6%), and these odds remained similar even after adjusting for other factors associated with election outcome: the unadjusted odds ratio (OR) was 4.25 (95% Confidence Intervals [CI]: 2.17-8.30), while the adjusted odds ratio (aOR) was 4.39 (95% CI, 1.79-10.81). However, this was the reverse for the MGB: the unadjusted and adjusted OR was significantly lower than one for the constituencies in the poorest district: 0.21 (95% CI, 0.10-0.41) and 0.26 (95% CI, 0.11-0.63). Similarly, the AIMIM factor reduced NDA's odds of winning an election, the unadjusted OR and aOR was 0.37 (95% CI, 0.14 to 1.01) and 0.09 (95% CI, 0.02-0.33), respectively. While for the MGB, the AIMIM factor improved the odds for winning an election, the OR and aOR was 0.99 (95% CI, 0.39-2.48) and 2.97 (95% CI, 0.91-9.73), respectively. The results of LJP factor are similar – it reduced NDA's odds of winning as reflected in OR and aOR of 0.37 (95% CI, 0.22-0.63) and 0.29 (95% CI, 0.15-0.54) and it increased MGB's odds of winning as reflected in OR and aOR of 2.80 (95% CI, 1.65-4.76) and 3.49 (95% CI, 1.87-6.49).

We also find that constituencies with higher sex ratio of the electorate voted in favor of the NDA, though this effect becomes lower and insignificant once we adjust for the proportion of the poorest households in the district in which the constituency is located. The OR and aOR were 2.24 (95% CI, 1.19-4.21) and 1.80 (95% CI, 0.78-4.13), respectively. While for the MGB it was the reverse, MGB was less likely to win in constituencies with higher sex ratio of the electorate.

4. **Discussion**

In this paper we explore the association between socio-economic factors and the likelihood of win at the level of the constituency. Our key finding is that NDA was more likely to win in constituencies that were the poorest. This is an important finding given that these elections were held in the midst of a pandemic where the poorest population were affected significantly (NCAER

COVID19 surveys, 2020). It seems to suggest that the different welfare and humanitarian schemes which were rolled out as immediate policy interventions after the lockdown were effective in reaching the poorest section of the population in Bihar.

Women votes have been important determinant of NDA win in Bihar assembly elections since 2005, but our analysis for 2020 elections reveals that once we control for poverty levels, then it is insignificant at the conventional 5% level. Hence, the results reinforce poverty as the key driver of election outcomes in Bihar 2020.

Another striking finding in this study, which goes against the popular narrative, is of Muslim population's voting preferences. The results strongly show that NDA was more likely to win in constituencies with higher proportion of Muslim population, for example, the aOR was 3.40 (95% CI, 1.24-9.34) for constituencies that were located in districts with Muslim population ≥15.33% as compared to the reference group, which were constituencies located in districts where the Muslim population <8.6%. Furthermore, the AIMIM and LJP factor significantly reduced the odds of NDA win in the constituency in which AIMIM and LJP contested the election, the aOR was 0.09 (95% CI, 0.02-0.33) and 0.29 (95% CI, 0.15-0.54), respectively, as compared to those constituencies in which they did not contest an election.

Post elections, there has been much noise on impact of AIMIM on performance of MGB with the popular belief that AIMIM played the saboteur to the INC in particular. Our analysis refutes this consistently because the 'AIMIM factor' improved the odds of MGB winning significantly, as reflected in the aOR of 2.97. Within the MGB, the AIMIM factor seems to have benefitted the INC in particular, as shown clearly in Figure 1b. The odds of INC winning are significantly higher in constituencies where the AIMIM contested than elsewhere. There is no such clear relationship for the other MGB partner, the RJD.

5. Limitations

First and foremost, while this paper establishes a strong relationship between socio-economic factors and election outcomes but it does not indicate neat causality due to numerous other observable and unobservable factors which might determine election outcomes. Second, we do not have data on socio-economic factors at the level of the constituency, which is only available at the level of a district. There are 243 assembly constituencies across 38 districts. On an average there are 6 to 7 assembly constituencies in each district, however, there is one district Sheohar, which has one assembly constituency, while, Patna district has 14 assembly constituencies. Third, our results are suggestive of factors that were associated with a win in a constituency, there may be other factors that are excluded from the analysis, which might affect the results.

6. **Conclusion**

Our paper suggests that in Bihar assembly elections 2020, poverty was the key driving factor, over and above female voters. The results show that the poor were more likely to support the NDA. The relevance of this result for an election held in the midst of a pandemic, is very crucial. It seems to suggest that even though the poor were the hardest hit by the COVID–19, the central and the state schemes for the benefit of the poor might have been an important factor in NDA's victory. Secondly, in contrast to conventional commentary, the AIMIM factor and the LJP factor hurt the NDA much more, in contrast, the MGB benefitted from their presence in the Bihar assembly elections.

**Table 1:** Summary of the results

|  | Total seats contested | Total seats won | Strike rate | Total votes | Votes (%) Overall | Votes (%) across constituencies contested |
|---|---|---|---|---|---|---|
| **Overall** | 243 |  |  | 42,137,620 |  |  |
| **NDA** | 245* | 125 | 51.4% | 15,701,226 | 37.3% |  |
| JDU | 115 | 43 | 37.4% | 6,484,414 | 15.4% | 32.8% |
| BJP | 110 | 74 | 67.3% | 8,201,408 | 19.5% | 42.6% |
| VIP | 13 | 4 | 30.8% | 639,840 | 1.5% | 27.7% |
| HAMS | 7 | 4 | 57.1% | 377,564 | 0.9% | 32.3% |
|  |  |  |  |  |  |  |
| **MGB** | 243 | 110 | 45.3% | 15,688,458 | 37.2% |  |
| RJD | 144 | 75 | 52.1% | 9,736,242 | 23.1% | 38.9% |
| INC | 70 | 19 | 27.1% | 3,995,003 | 9.5% | 32.9% |
| CPI (ML) (L) | 19 | 12 | 63.2% | 1,333,569 | 3.2% | 41.4% |
| CPI | 6 | 2 | 33.3% | 349,489 | 0.8% | 33.3% |
| CPM | 4 | 2 | 50.0% | 274,155 | 0.7% | 37.6% |

---

* In constituencies Sikti and Kishanganj, BJP and VIP, who were a part of the NDA alliance fielded a candidate.

**Table 2a:** Summary statistics at the constituency level based on the district level data from NFHS IV

|  | Total districts | Total constituencies | (%) |
|---|---|---|---|
| ***Sex ratio of the electorate (female per 1000 male electors)*** | | | |
| < 861 |  | 81 | 33.3% |
| [861 to 886) |  | 81 | 33.3% |
| ≥ 886 |  | 81 | 33.3% |
| ***Proportion of poorest household*** | | | |
| < 15.5% | 13 | 78 | 32.1% |
| [15.5% to 25.5%) | 13 | 87 | 35.8% |
| ≥ 25.5% | 12 | 78 | 32.1% |
| ***Proportion of Muslim population*** | | | |
| < 8.6% | 13 | 71 | 29.2% |
| [8.6% to 15.33%) | 13 | 86 | 35.4% |
| ≥ 15.33% | 12 | 86 | 35.4% |
| ***Proportion of general caste*** | | | |
| < 14.5% | 13 | 69 | 28.4% |
| [14.5% to 19.6%) | 13 | 94 | 38.7% |
| ≥ 19.6% | 12 | 80 | 32.9% |
| ***Proportion of population in rural areas*** | | | |
| < 86.6% | 13 | 85 | 35.0% |
| [86.6% to 92.3%) | 13 | 87 | 35.8% |
| ≥ 92.3% | 12 | 71 | 29.2% |
| ***AIMIM factor*** | | | |
| Did not contest |  | 223 | 91.8% |
| Contested |  | 20 | 8.2% |
| ***LJP factor*** | | | |
| Did not contest |  | 108 | 44.4% |
| Contested |  | 135 | 55.6% |

**Table 2b:**

| | NDA Seats contested, won (strike rate) | JDU Seats contested, won (strike rate) | BJP Seats contested, won (strike rate) | MGB Seats contested, won (strike rate) | RJD Seats contested, won (strike rate) | INC Seats contested, won (strike rate) |
|---|---|---|---|---|---|---|
| *Sex ratio of the electorate (female per 1000 male electors)* | | | | | | |
| < 861 | 81  36 (44.4%) | 31  5 (16.1%) | 45  30 (66.7%) | 81  44 (54.3%) | 48  30 (62.5%) | 22  6 (27.3%) |
| [861 to 886) | 81  37 (45.7%) | 45  15 (33.3%) | 31  20 (64.5%) | 81  39 (48.1%) | 48  25 (52.1%) | 22  8 (36.4%) |
| ≥ 886 | 81  52 (64.2%) | 39  23 (59.0%) | 34  24 (70.6%) | 81  27 (33.3%) | 48  20 (41.7%) | 26  5 (19.2%) |
| *Proportion of poorest household* | | | | | | |
| < 15.5% | 78  27 (34.6%) | 39  10 (25.6%) | 37  17 (45.9%) | 78  50 (64.1%) | 43  32 (74.4%) | 23  10 (43.5%) |
| [15.5% to 25.5%) | 87  44 (50.6%) | 39  12 (30.8%) | 36  24 (66.7%) | 87  39 (44.8%) | 55  29 (52.7%) | 22  5 (22.7%) |
| ≥ 25.5% | 78  54 (69.2%) | 37  21 (56.8%) | 37  33 (89.2%) | 78  21 (26.9%) | 46  14 (30.4%) | 25  4 (16.0%) |
| *Proportion of Muslim population* | | | | | | |
| < 8.6% | 71  26 (36.6%) | 41  11 (26.8%) | 24  12 (50.0%) | 71  45 (63.4%) | 42  30 (71.4%) | 19  8 (42.1%) |
| [8.6% to 15.33%) | 86  41 (47.7%) | 39  13 (33.3%) | 42  25 (59.5%) | 86  42 (48.8%) | 54  32 (59.3%) | 21  5 (23.8%) |
| ≥ 15.33% | 86  58 (67.4%) | 35  19 (54.3%) | 44  37 (84.1%) | 86  23 (26.7%) | 48  13 (27.1%) | 30  6 (20.0%) |
| *Proportion of general caste* | | | | | | |
| < 14.5% | 69  41 (59.4%) | 41  20 (48.8%) | 27  20 (74.1%) | 69  26 (37.7%) | 40  16 (40.0%) | 26  7 (26.9%) |
| [14.5% to 19.6%) | 94  46 (48.9%) | 42  14 (33.3%) | 44  27 (61.4%) | 94  45 (47.9%) | 60  36 (60.0%) | 19  3 (15.8%) |
| ≥ 19.6% | 80  38 (47.5%) | 32  9 (28.1%) | 39  27 (69.2%) | 80  39 (48.8%) | 44  23 (52.3%) | 25  9 (36.0%) |
| *Proportion of population in rural areas* | | | | | | |
| < 86.6% | 85  41 (48.2%) | 40  12 (30.0%) | 40  26 (65.0%) | 85  41 (48.2%) | 42  26 (61.9%) | 29  6 (20.7%) |
| [86.6% to 92.3%) | 87  42 (48.3%) | 35  10 (28.6%) | 40  27 (67.5%) | 87  42 (48.3%) | 56  28 (50.0%) | 25  10 (40.0%) |
| ≥ 92.3% | 71  42 (59.2%) | 40  21 (52.5%) | 30  21 (70.0%) | 71  27 (38.0%) | 46  21 (45.7%) | 16  3 (18.8%) |
| *AIMIM factor* | | | | | | |
| Did not contest | 223  119 (53.4%) | 104  41 (39.4%) | 104  71 (68.3%) | 223  101 (45.3%) | 133  72 (54.1%) | 63  15 (23.8%) |
| Contested | 20  6 (30.0%) | 11  2 (18.2%) | 6  3 (50.0%) | 20  9 (45.0%) | 11  3 (27.3%) | 7  4 (57.1%) |
| *LJP factor* | | | | | | |
| Did not contest | 108  70 (64.8%) | 2  0 (0.0%) | 104  70 (67.3%) | 108  34 (31.5%) | 61  21 (34.4%) | 33  6 (18.2%) |
| Contested | 135  55 (40.7%) | 113  43 (38.1%) | 6  4 (66.7%) | 135  76 (56.3%) | 83  54 (65.1%) | 37  13 (35.1%) |

Note: For each alliance or the party the first column is the total number of seats contested in the constituency with the characteristic, the second column are the seats won, and the percentage in brackets is numbers of seats won as compared to the number of seats contested.

**Table 3a: Odds ratios for NDA**

|  | Unadjusted | | Adjusted | |
|---|---|---|---|---|
|  | **Odds ratio** | **95% CI** | **Odds ratio** | **95% CI** |
| *Sex ratio of the electorate (female per 1000 male electors)* | | | | |
| < 861 | 1* | | 1* | |
| [861 to 886) | 1.05 | (0.57 to 1.95) | 1.01 | (0.46 to 2.21) |
| ≥ 886 | 2.24 | (1.19 to 4.21) | 1.80 | (0.78 to 4.13) |
| *Proportion of poorest household* | | | | |
| < 15.5% | 1* | | 1* | |
| [15.5% to 25.5%) | 1.93 | (1.03 to 3.62) | 2.97 | (1.22 to 7.21) |
| ≥ 25.5% | 4.25 | (2.17 to 8.30) | 4.39 | (1.79 to 10.81) |
| *Proportion of Muslim population* | | | | |
| < 8.6% | 1* | | 1* | |
| [8.6% to 15.33%) | 1.58 | (0.83 to 3.00) | 1.04 | (0.44 to 2.44) |
| ≥ 15.33% | 3.59 | (1.85 to 6.94) | 3.40 | (1.24 to 9.34) |
| *Proportion of general caste* | | | | |
| < 14.5% | 1* | | 1* | |
| [14.5% to 19.6%) | 0.65 | (0.35 to 1.23) | 0.40 | (0.19 to 0.87) |
| ≥ 19.6% | 0.62 | (0.32 to 1.18) | 0.35 | (0.14 to 0.85) |
| *Proportion of population in rural areas* | | | | |
| < 86.6% | 1* | | 1* | |
| [86.6% to 92.3%) | 1.00 | (0.55 to 1.82) | 0.58 | (0.26 to 1.30) |
| ≥ 92.3% | 1.55 | (0.82 to 2.94) | 0.68 | (0.26 to 1.83) |
| *AIMIM factor* | | | | |
| Did not contest | 1* | | 1* | |
| Contested | 0.37 | (0.14 to 1.01) | 0.09 | (0.02 to 0.33) |
| *LJP factor* | | | | |
| Did not contest | 1* | | 1* | |
| Contested | 0.37 | (0.22 to 0.63) | 0.29 | (0.15 to 0.54) |

**Table 3b:** Odds ratio for MGB

|  | Unadjusted | | Adjusted | |
| --- | --- | --- | --- | --- |
|  | Odds ratio | 95% CI | Odds ratio | 95% CI |
| *Sex ratio of the electorate (female per 1000 male electors)* | | | | |
| < 861 | 1* | | 1* | |
| [861 to 886) | 0.78 | (0.42 to 1.45) | 0.89 | (0.41 to 1.93) |
| ≥ 886 | 0.42 | (0.22 to 0.79) | 0.60 | (0.26 to 1.38) |
| *Proportion of poorest household* | | | | |
| < 15.5% | 1* | | 1* | |
| [15.5% to 25.5%) | 0.46 | (0.24 to 0.85) | 0.33 | (0.13 to 0.79) |
| ≥ 25.5% | 0.21 | (0.10 to 0.41) | 0.26 | (0.11 to 0.63) |
| *Proportion of Muslim population* | | | | |
| < 8.6% | 1* | | 1* | |
| [8.6% to 15.33%) | 0.55 | (0.29 to 1.05) | 0.79 | (0.34 to 1.85) |
| ≥ 15.33% | 0.21 | (0.11 to 0.42) | 0.23 | (0.08 to 0.65) |
| *Proportion of general caste* | | | | |
| < 14.5% | 1* | | 1* | |
| [14.5% to 19.6%) | 1.52 | (0.81 to 2.86) | 2.38 | (1.10 to 5.15) |
| ≥ 19.6% | 1.57 | (0.82 to 3.03) | 3.03 | (1.24 to 7.43) |
| *Proportion of population in rural areas* | | | | |
| < 86.6% | 1* | | 1* | |
| [86.6% to 92.3%) | 1.00 | (0.55 to 1.82) | 1.91 | (0.85 to 4.27) |
| ≥ 92.3% | 0.66 | (0.35 to 1.25) | 1.67 | (0.62 to 4.54) |
| *AIMIM factor* | | | | |
| Did not contest | 1* | | 1* | |
| Contested | 0.99 | (0.39 to 2.48) | 2.97 | (0.91 to 9.73) |
| *LJP factor* | | | | |
| Did not contest | 1* | | 1* | |
| Contested | 2.80 | (1.65 to 4.76) | 3.49 | (1.87 to 6.49) |

*Reference group for the analysis.

**Figure 1a:** Unadjusted and Adjusted odds ratios for NDA and MGB

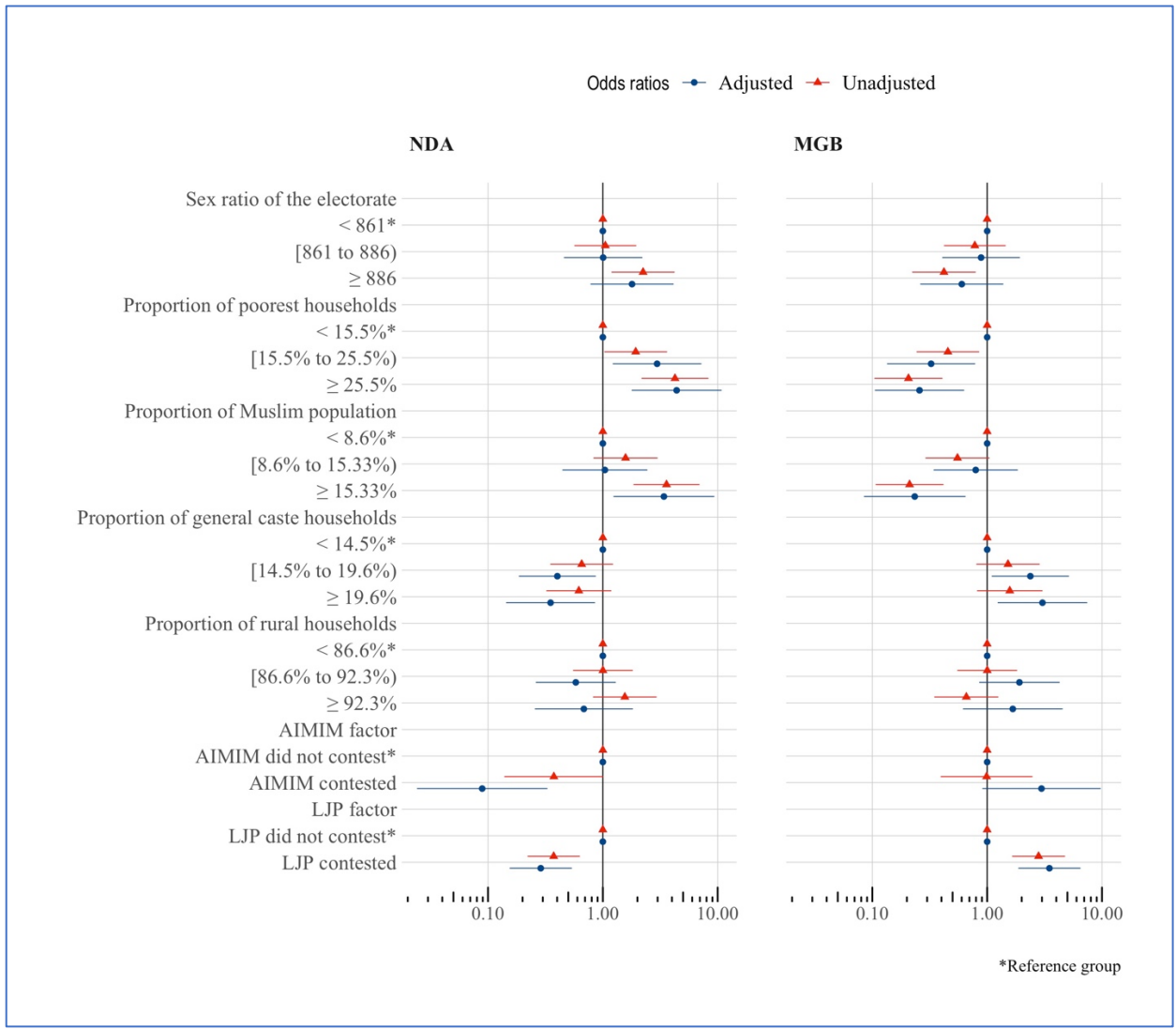

**Figure 1b:** Unadjusted and Adjusted odds ratios for JDU, BJP, RJD, INC

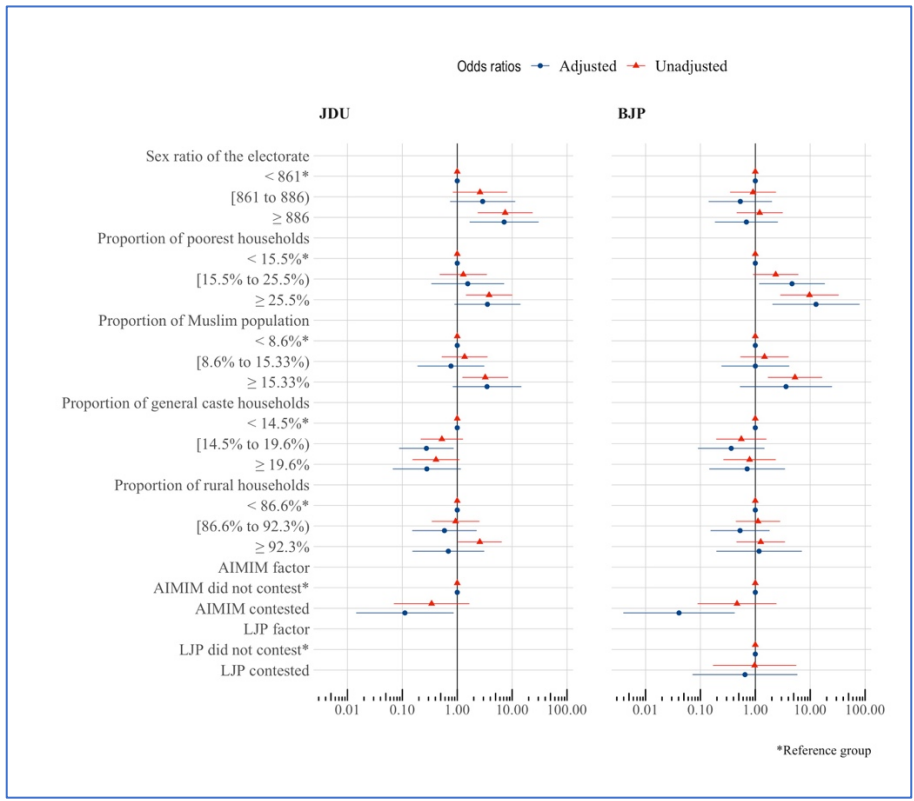

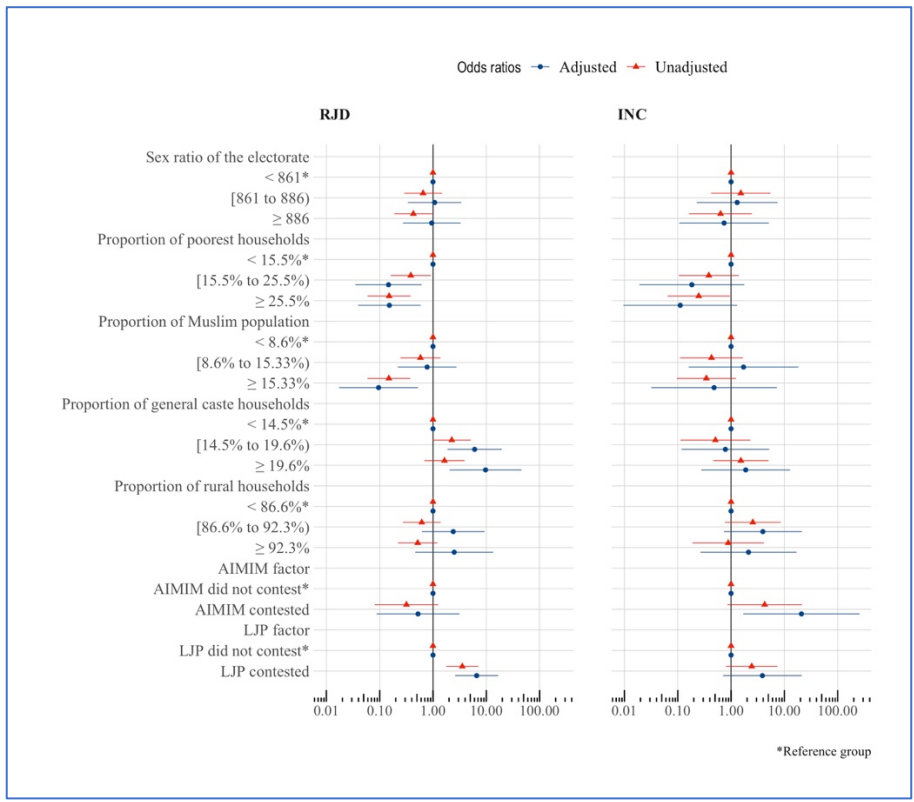